\documentclass[12pt]{article}

\usepackage[T1]{fontenc}         % Improved font encoding
\usepackage[utf8]{inputenc}        % UTF-8 input encoding
\usepackage{lmodern}               % Enhanced font rendering
\usepackage{graphicx} 
\usepackage{authblk}
\usepackage{amssymb}
\usepackage{amsmath}
\usepackage{underscore}

\title{A tertiary review on quantum cryptography}

\author{Luiz Filipe Anderson de Sousa Moura\\
Carlos Becker Westphall\\
\thanks{luiz.f.s.m@posgrad.ufsc.br} \quad \thanks{carlos.westphall@ufsc.br}\\
Department of Informatics and Statistics\\
Federal University of Santa Catarina\\
Florianopolis\\
Brazil}

\begin{document}

\maketitle

%% Abstract
\begin{abstract}
Quantum computers impose an immense threat to system security. As a countermeasure, new cryptographic classes have been created to prevent these attacks. Technologies such as post-quantum cryptography and quantum cryptography. Quantum cryptography uses the principle of quantum physics to produce theoretically unbreakable security. This tertiary review selected 51 secondary studies from the Scopus database and presented bibliometric analysis, a list of the main techniques used in the field, and existing open challenges and future directions in quantum cryptography research. The results showed a prevalence of QKD over other techniques among the selected papers and stated that the field still faces many problems related to implementation cost, error correction, decoherence, key rates, communication distance, and quantum hacking.
\end{abstract}

%% Keywords

\section{Introduction}
Quantum computers leverage the principles of quantum mechanics to produce advanced computations, being theoretically able to solve a variety of computational tasks considered infeasible for classical computers. These tasks include a great part of the NP class of problems, revealing opportunities for future research in different areas, but also imposing threats to the security of computational systems as we understand them. 

To fight against this threat, Post-Quantum Cryptography (PQC) and new techniques of quantum cryptography were created. PQC algorithms are designed to be secure against classical and quantum attacks and to operate on classical hardware. These algorithms rely tipically on mathematical problems considered hard to solve, even for quantum computers. Quantum cryptography, on the other hand, is a type of cryptography that wields the natural laws of quantum physics to produce theoretically insurmountable security and require the implementation of special hardware.

The object of this research is to provide an overview of the state-of-the-art of the field, deepening the knowledge, but also pointing gaps for further research. More specifically, the goals of this study can be stated as: i) to map existing secondary reviews on quantum cryptography, ii) to map different quantum cryptography techniques, and iii) to find tendencies and gaps in the field of study.

A tertiary literature review is a special case of systematic study whose object is to synthesize results from existing secondary studies (i.e. systematic studies or surveys) in order to achieve a deeper understanding of the state-of-the-art or even to ask questions that were not possible in a primary or secondary study. To better complete these objectives, this study will follow the research protocol presented in Section 2. Then, an analysis of the selected papers will be presented in Section 3. Section 4 makes evident the points where this research was not perfect. Section 5 presents relevant keywords in quantum cryptography techniques. Section 6 shows trends, open challenges, and existing gaps in quantum cryptography. And Section 7 is the conclusion.

\section{Methods}

The present work is motivated by the current advancements in the field of quantum cryptography. To guide this tertiary review and help achieve the goals presented in the first section, the research protocol presented in Table~\ref{tab:protocol} was created.

\begin{table}[!htb]
    \centering
    \caption{Research protocol}
    \label{tab:protocol}
    \begin{tabular}{ll}
    \hline
        1. Research questions:  & RQ1) What are the selected studies?\\
                                & RQ2) What are the different quantum\\
                                & cryptography techniques?\\
                                & RQ3) What can we conclude about the\\
                                & tendencies and existing gaps in the field?\\
        2. Database:            & Scopus\\
        3. Search criteria:     & Research papers from the last twelve years.\\
                                & The papers must also include, in the title,\\
                                & abstract, or keywords, the terms “quantum\\
                                & cryptography” or “quantum key cryptography”\\
                                & or “quantum key distribution”, and also the\\
                                & terms “survey” or “systematic review” or\\
                                & “systematic mapping”.\\
                                & The exact search string: TITLE-ABS-KEY(\\
                                & (\{quantum cryptography\} OR \{quantum key\\
                                & cryptography\} OR \{quantum key\\
                                & distribution\}) AND (survey OR \{systematic\\
                                & review\} OR \{systematic mapping\})) AND\\
                                & PUBYEAR $>$ 2013 AND PUBYEAR $<$ 2026\\
        4. Screening:           & 258 were found using the search terms.\\
                                & Search made on March 3rd, 2025.\\
        5. Exclusion criteria:  & EC1) The paper is not a secondary study.\\
                                & EC2) The paper is not in the field of\\
                                & quantum cryptography.\\
        6. Selection process:   & The selection will be made in two steps:\\
                                & i) analyzing the abstract, ii) full text.\\
    \hline
    \end{tabular}
\end{table}

\subsection{The selection process}
This study used the Scopus\footnote{https://www.scopus.com} database for initial research. This decision was based on the wide range of peer-reviewed academic material available on the platform and the convenient search features. 

After applying the search criteria in the selected database, the next step is to perform the selection process to judge which papers should be included. In this work, the selection process was conducted in two levels: i) applying the exclusion criteria only in the abstract, and ii) applying the exclusion criteria in the full text.

\subsection{Quality score}
After the final stage of the selection process, the papers received a quality score and an affinity score. These scores were used to evaluate the general quality of the selected papers, the most relevant ones, and to gather additional information that might be relevant. The evaluation criteria were based on:
\begin{itemize}
    \item \textbf{Level of correlation to the field (C):} A strong correlation received 1 point, a moderate correlation received 0.5 points, and papers loosely related were automatically excluded.
    \item \textbf{Being a systematic review (S):} Being a tertiary review, only surveys or systematic studies were accepted. But systematic studies received 1 point in the quality score, surveys received 0 points, and semi-systematic studies (studies with a systematized search criteria, but not a full research protocol) received 0.5 points.
    \item \textbf{Year of publication (Y):} Papers published in 2025 received 1 score point. If published in 2014, 0 points. Other publication years received $\frac{YEAR - 2014}{11}$ points.
    \item \textbf{Number of citations (N):} The paper with the highest number of citations (MNC) received 1 point. Papers with a different number of citations (NC) received $\frac{NC}{MNC}$ points.
\end{itemize}

The quality score was then calculated, adding $C + S + Y + N$. It is also important to note that Y and N could sum up to two if a paper was published in 2025 and had $NC = MNC$. However, this is improbable and $Y + N > 1$ is expected to happen rarely.

The affinity score was calculated considering the highest quality score possible, four, defined as a percentage $\frac{QUALITY\_SCORE}{4}100\%$.

\subsection{Data extraction}
With the final list of selected papers, the data extraction procedure listed the most frequently mentioned techniques in quantum cryptography, eliminating techniques that appeared in fewer than three different materials. The procedure for finding open challenges in the field required a precise reading of the selected texts and depended on the author's experience in discriminating the most relevant information.

\section{Analysis of the selected papers}
\begin{figure}[!htb]
    \centering
    \includegraphics[width=1\linewidth]{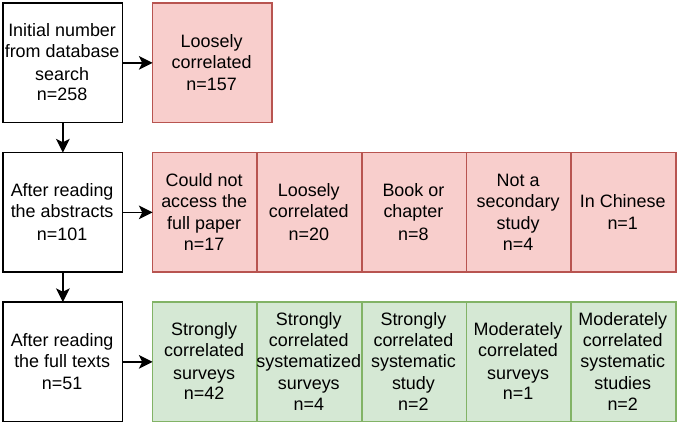}
    \caption{Selection process}
    \label{fig:selection}
\end{figure}

From the 258 initial papers found after the search in the database, 101 remained after reading the abstracts and 51 were selected after reading the full texts, as shown in Figure~\ref{fig:selection}.

157 papers were considered loosely correlated with the field after reading the abstracts. After reading the full texts, 20 were considered loosely correlated with the field, 4 were not secondary studies, 1 was in Chinese, 8 were books or book chapters, and 17 were not accessible through institutional login. A total of 50 papers or materials were excluded in this step.

Among the included papers, the majority were considered strongly correlated to the field (48 out of 51), as shown by the average C of 0.971. On the other hand, very few were systematic studies or systematized surveys (8 out of 51). The effect is made evident by the low average S of 0.118. Table~\ref{tab:papers} displays the selected papers with respective affinity scores.

\begin{table}[!htb]
    \centering
    \caption{Selected papers}
    \label{tab:papers}
    \begin{tabular}{llll}
    \hline
        \textbf{Reference} & \textbf{Affinity score} & \textbf{Reference} & \textbf{Affinity score}\\
    \hline
        \cite{b1}  & 72.75\% &	\cite{b27} & 44.01\%\\
        \cite{b2}  & 70.45\% &	\cite{b28} & 43.72\%\\
        \cite{b3}  & 59.04\% &	\cite{b29} & 43.66\%\\
        \cite{b4}  & 58.58\% &	\cite{b30} & 43.36\%\\
        \cite{b5}  & 58.01\% &	\cite{b31} & 42.43\%\\
        \cite{b6}  & 57.95\% &	\cite{b32} & 41.25\%\\
        \cite{b7}  & 54.42\% &	\cite{b33} & 41.12\%\\
        \cite{b8}  & 53.62\% &	\cite{b34} & 39.83\%\\
        \cite{b9}  & 50.00\% &	\cite{b35} & 39.47\%\\
        \cite{b10} & 50.00\% &	\cite{b36} & 37.76\%\\
        \cite{b11} & 48.14\% &	\cite{b37} & 37.71\%\\
        \cite{b12} & 47.96\% &	\cite{b38} & 36.80\%\\
        \cite{b13} & 47.83\% &	\cite{b39} & 36.39\%\\
        \cite{b14} & 47.73\% &	\cite{b40} & 35.67\%\\
        \cite{b15} & 47.73\% &	\cite{b41} & 35.38\%\\
        \cite{b16} & 47.73\% &	\cite{b42} & 35.31\%\\
        \cite{b17} & 47.73\% &	\cite{b43} & 34.92\%\\
        \cite{b18} & 47.73\% &	\cite{b44} & 34.53\%\\
        \cite{b19} & 46.54\% &	\cite{b45} & 33.40\%\\
        \cite{b20} & 45.51\% &	\cite{b46} & 32.34\%\\
        \cite{b21} & 45.48\% &	\cite{b47} & 30.71\%\\
        \cite{b22} & 45.45\% &	\cite{b48} & 29.70\%\\
        \cite{b23} & 44.48\% &	\cite{b49} & 28.62\%\\
        \cite{b24} & 44.45\% &	\cite{b50} & 27.30\%\\
        \cite{b25} & 44.35\% &	\cite{b51} & 25.10\%\\
        \cite{b26} & 44.19\%\\	
    \hline
    \end{tabular}
\end{table}

The average affinity score of the papers was 43.888\% (average quality score of 1.755) and only two had $Y + N \ge 1$. Most papers were strongly correlated with the field, but the number of systematic studies was low. This was the main factor that lowered the affinity score.

\begin{figure}[!htb]
    \centering
    \includegraphics[width=1\linewidth]{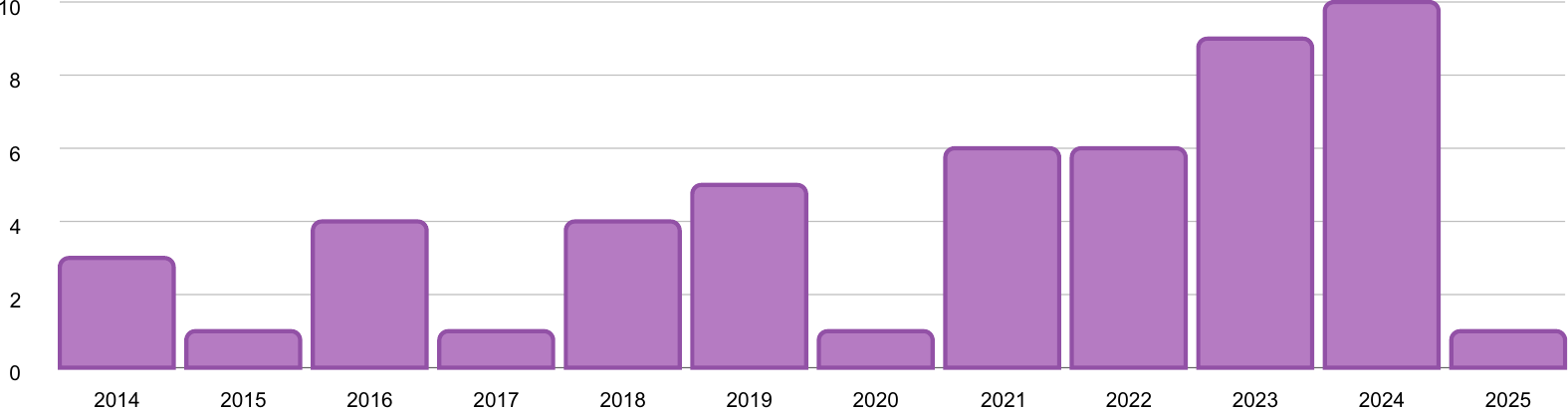}
    \caption{Selection process}
    \label{fig:years}
\end{figure}

The average paper had 20.510 pages and an average number of 57.490 citations per paper. The median year of publication was 2022. Figure~\ref{fig:years} is a histogram showing the number of papers that fall within January and December of each year, from 2014 to 2025 (2025 was considered from January to March). In this figure, it is possible to see the increasing number of papers that fall within the quantum cryptography field.

Of the selected papers, 22 were articles, 14 were conference papers, 13 were review papers, and only one was considered a short survey.

It is also important to mention that no other tertiary systematic review was found in this research, even if it would not be included in the study, according to the exclusion criteria.

\section{Threats to Validity}

Despite following a systematic and transparent research protocol, the research is not free from threats to validity that must be acknowledged:

\begin{itemize}
    \item \textbf{Selection bias:} The initial search used exclusively the Scopus database and, despite having a large number of indexed material, relevant studies might be outside the scope. Time and language constraints can also limit the scope for relevant studies, but might be less relevant, since most studies should fall within these criteria.
    \item \textbf{Screening and score subjectivity:} The screening process divided into two steps, first reading the abstracts and later the full texts, was carried out by a researcher, and it is not free of human errors. The quality score, although standardized, inherit subjectivity in its formulation.
    \item \textbf{Data Extraction Limitations:} During the data extraction, every abstract were available for classification. However, when reading the full texts, 17 out of 101 papers ($\approx 16.832\%$) could not be accessed through institutional login. Also, the classification of strong, moderate, or loosely correlated is subjective and is based on the experience of the authors.
    \item \textbf{Generalizability:} This research was based on secondary studies and since the number of systematic or systematized studies was low (8 out of 51), so is the understanding of the protocol used in the studies considered and lower is the generalizability. For this reason, the quality score is important, and materials with a higher score were considered more relevant.
\end{itemize}

Although every effort was made to minimize these threats and the protocol is transparent, this review can be better interpreted with these limitations in mind.

\section{The quantum cryptography techniques}

\begin{figure}[!htb]
    \centering
    \includegraphics[width=1\linewidth]{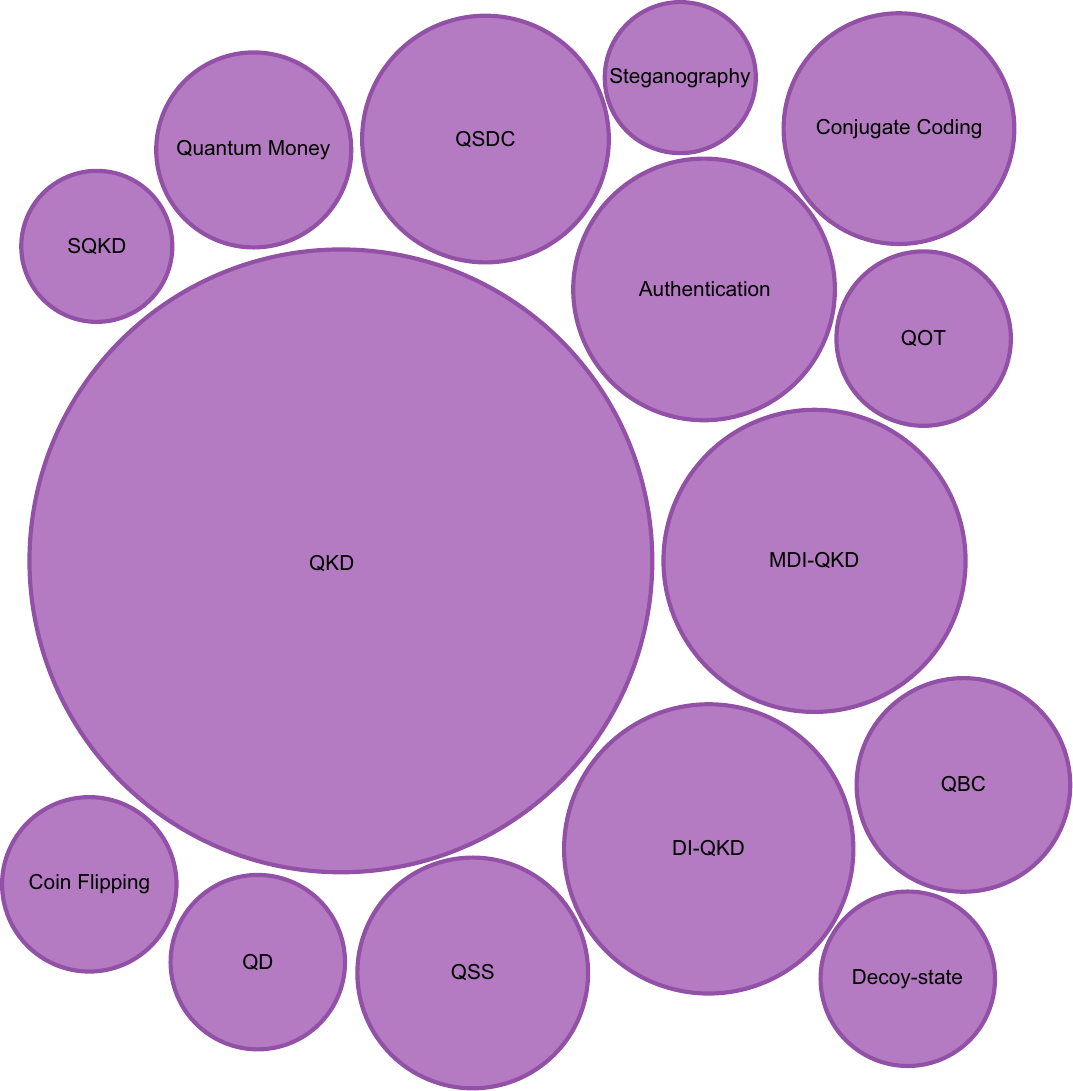}
    \caption{Main keywords in quantum cryptography techniques}
    \label{fig:techniques}
\end{figure}

This study mapped different quantum cryptography techniques. Figure~\ref{fig:techniques} lists the main ones with respective frequency.

The following text summarizes the techniques listed in Figure~\ref{fig:techniques}.

\subsection{Summary of techniques}

The term Quantum Key Distribution (QKD) appeared in every selected paper and, hence, is the most popular technique by a large margin. According to~\cite{b31}, there are two schemes of QKD protocols: one based on Heisenberg's uncertainty principle and on the no-cloning theorem, the Prepare \& Measure (P\&M) scheme; and the Entanglement-Based (EB) scheme. Also, according to~\cite{b31}, there are three families of QKD protocols: Discrete-Variable (DV), Continuous-Variable (CV), and Distributed-Phase Reference (DPR) QKD. Here is a short explanation of each scheme and family:
\begin{itemize}
    \item \textbf{P\&M:} A prepares and sends polarized photons to B.
    \item \textbf{EB:} A source generates pairs of entangled photons and sends them to both A and B.
    \item \textbf{DV-QKD:} Uses discrete parameters, such as the polarization of photons, to encode.
    \item \textbf{CV-QKD:} Encodes using continuous parameters, such as the electromagnetic field of light, amplitude, or phase.
    \item \textbf{DPR-QKD:} A and B agree on a reference phase distributed along with the key signals. This reduces problems related to phase desynchronization.
\end{itemize}

Measurement-Device-Independent QKD (MDI-QKD) is a class of QKD protocols independent of the measurement device (the detector). This class of protocols is interesting because it decreases the chances of a key distribution being hacked by side-channel attacks. Similarly, Device-Independent QKD (DI-QKD) is another class of QKD protocols committed to ensure that the key is established independent of the devices, from both sides, generation and detection.

Following the list, quantum authentication and quantum signature were brought together because they have similar applications. Authentication is the process of guaranteeing that someone is exactly who they say. A signature is a method of proving one's identity. Instead of leveraging mathematical methods, quantum authentication/signature relies on quantum mechanics to perform their respective tasks.

Instead of establishing a secret key before message exchange, Quantum Secure Direct Communications (QSDC) encodes the message directly into quantum states and transfers them through the quantum channel.

The secret sharing problem tries to solve a situation where a secret has to be distributed, but in such a way that no individual can gather information about the secret. However, when a sufficient number of individuals join their shares, the secret can be revealed. Classical secret sharing frequently uses mathematics or image manipulation to accomplish this task, and Quantum Secret Sharing (QSS) tries to use quantum properties.

Conjugate coding is a concept that is older than quantum computing itself. It refers to the use of conjugate bases to encode information in quantum cryptography. This primitive was used in many quantum cryptographic protocols.

Quantum money is also a concept older than quantum computing itself. The concept leverages quantum mechanics to create a currency with unforgeability and easier fraud detection.

Quantum Bit Commitment (QBC) or classical bit commitment work the following way: (1) A encodes a bit (0 or 1) and sends it to B, with the properties that B cannot see the bit until A reveals it and A cannot change the committed bit; (2) A reveals the bit and proves that the bit was not modified. Unconditional security for QBC is considered impossible, so security measurements should be included in protocols using this primitive~\cite{b33}.

Quantum Dialogue (QD) distinguishes itself from other quantum cryptographic techniques for allowing bidirectional communication. Whilst most quantum cryptographic methods are focused on transmitting a single message, QD focus on two parties exchanging messages simultaneously.

Quantum Oblivious Transfer (QOT) implements the classical Oblivious Transfer (OT) in the quantum realm. In OT, A sends $n$ messages ($m_0$, $m_1$, ..., $m_n$) to B, who chooses one of the messages and A remains oblivious about B's choice. In the case of an arbitrary $n$, the process is called 1-out-of-n Oblivious Transfer, and in the special case of $n = 2$, it is called 1-out-of-2 Oblivious Transfer.

Semi-Quantum Key Distribution (SQKD), is a class of key distribution protocols that embraces the problem of most devices not having quantum capabilities and performing the key agreement between an end with quantum capabilities (usually the sender) and an end with limited quantum capabilities (usually the receiver).

Quantum steganography is another self-explanatory concept. Similarly to classical steganography, quantum steganography is the practice or the art of hiding secret information in quantum states in such a way that no other person (or unauthorized people) could notice there is hidden information.

In many cryptographic protocols, A and B trust each other, but the threat comes from the outside (insecure network, unauthorized people, etc.). The coin flipping problem considers that even A and B do not trust each other. The problem is that A and B have to agree on a random bit only by exchanging messages and without any other trusted device. In classical coin flipping, one of the parties (A or B) may manipulate the outcome. Quantum coin flipping reduces this threat by leveraging quantum mechanical principles.

Decoy-state is a technique specially used to reduce the susceptibility of QKD protocols to the Photon-Number-Splitting (PNS) attack by transmitting pulses in random intensity levels. With this, the possibility that a hacker could split the light pulse and remain undetected is avoided.

\subsection{Summary of keywords in QKD}

\begin{figure}[!htb]
    \centering
    \includegraphics[width=1\linewidth]{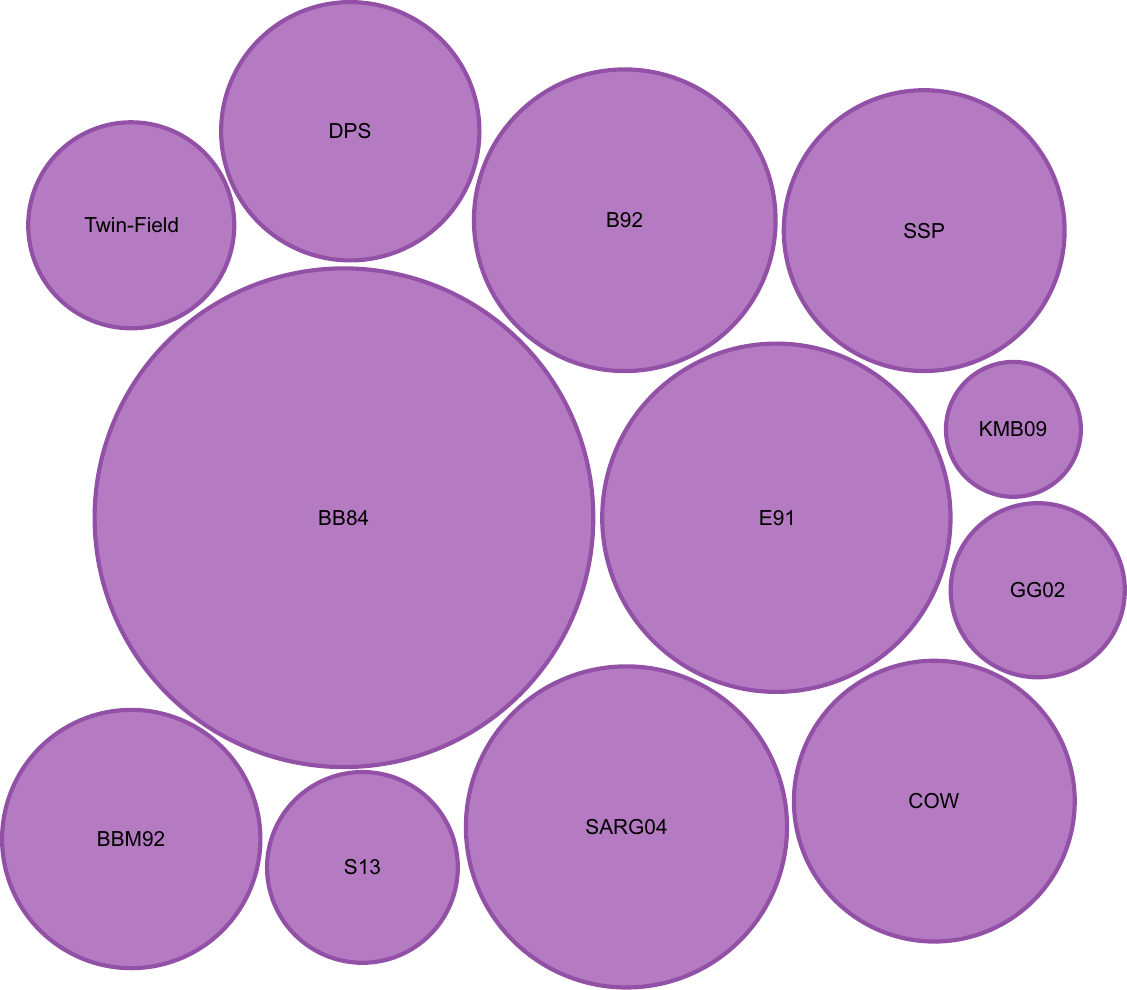}
    \caption{Main keywords in QKD protocols}
    \label{fig:qkdprotocols}
\end{figure}

As seen previously, QKD is the most popular quantum cryptography technique. Figure~\ref{fig:qkdprotocols} lists the most mentioned QKD protocols in this study with their respective frequency. This section will also briefly present each one of the protocols. 

BB84~\cite{b52} is known as the first QKD protocol, released in 1984. This is a P\&M protocol that works with discrete-variables. The protocol was inspired by conjugate coding and utilizes conjugate bases (rectilinear and diagonal) to encode the qubits. The protocol can be explained shortly as: (1) A chooses random basis for each bit and sends the photons to B, (2) B randomly chooses the bases to measure each qubit, (3) they share publicly their chosen bases and keep only the bits that match, (4) these bits form the "raw key", (5) they investigate on the raw key for possible error or eavesdropping during transmission, (6) the raw key is refined with classical methods into a final communication key.

E91~\cite{b53} protocol, sometimes called Eckert's protocol, is one of the first QKD protocols. It is a DV-QKD protocol based on quantum entanglement. A source generates a pair of entangled photons and sends them to A and B. A and B then measure them with random basis. Their measurements are correlated as a consequence of the entanglement.

SARG04~\cite{b54} is a variation of BB84 presented previously. This protocol is more robust than BB84 and is known for being more resistant to PNS attacks. By transmitting a lower level light pulse (say, one or fewer photons per pulse), SARG04 can possibly avoid that a hacker split the light pulse, storing the remaining in a quantum memory.

B92~\cite{b55} is another variation of the BB84 protocol. This protocol, proposed by Benet in 1992, uses two non-orthogonal quantum states instead of four. It can be understood as a simpler version of BB84.

SSP~\cite{b56} stands for Six-State Protocol. This protocol is also a variation of BB84, but using three conjugate bases to encode, instead of two. By using three bases, it uses a total of six quantum states to encode information. However, the higher number of states results in a slower performance. But the error detection caused by eavesdropping is considered higher.

The Coherent One-Way (COW)~\cite{b57} is a DPR-QKD protocol. It sends weak light pulses and relies on the coherence of consecutive pulses. The Differential Phase Shift (DPS)~\cite{b58} is also a DPR-QKD protocol, released one year earlier. This protocol also uses the relative phase between weak coherent light pulses to encode information.

BBM92~\cite{b59} is related to BB84, but it is an EB protocol. A source generates a pair of entangled photons and sends them to A and B. They then measure them with random bases. The conjugate bases used in this protocol are two: rectilinear and diagonal.

S13~\cite{b60} is a MDI-QKD protocol, which means it does not rely on the trust of the measurement device and is safer against side-channel attacks. It uses a third party for measurement, and the transmission is done with weak coherent pulses, also being safer against PNS attacks. This protocol uses entanglement, but also shares similarities with the P\&M scheme. 

Twin-Field~\cite{b61} protocol is designed to enable long distance communication. This protocol combines P\&M and EB schemes. With decoy-state technique, this protocol is also resistant to PNS attacks.

KMB09~\cite{b62} is an EB QKD protocol. It transmits weak coherent pulses and, just like S13, is more resistant to PNS attacks than other weak coherent pulse protocols for relying on quantum entanglement.

\section{Tendencies and existing gaps in quantum cryptography}

Quantum cryptography is a promising field of study in the interest of Physics, Computer Science, and Engineering. With an increasing number of publications in the last three decades and a market expected to be worth \$291.9 million by 2026 and \$455.3 millions by 2030, with USA and China controlling the largest market share~\cite{b1}.

\iffalse
A list of application areas where quantum cryptography is relevant is presented by~\cite{b1}:
\begin{itemize}
    \item Secure Communication.
    \item Financial Services.
    \item Data Centers and Cloud Environment.
    \item IoT Security.
    \item Healthcare and Medical Data.
    \item Authentication and Identity Management.
    \item Secure Election and Voting Systems.
\end{itemize}
\fi

According to~\cite{b2}, papers published before 2015 were generally focused on creating new protocols and studying the different quantum cryptography techniques (such as QKD, QSDC, QBC, QOT, Quantum signature, and Coin Tossing); whilst more recent papers have a general focus on combining different existing protocols to create new ones, further optimizing protocols, implementing them in different situations, or analyzing the limitations of existing protocols. The same paper concludes that Quantum Secret Sharing, Quantum Bit Commitment, and One-Time-Pad are keywords associated with older research papers; and Quantum Private Comparison, Deterministic Protocols, Quantum Entanglement, and Quantum Teleportation are keywords used more recently.

In conformity to what~\cite{b11} said, quantum computers will become operational eventually, it is just a matter of when. And current telecommunications around the world could be considered insecure if the quantum threat becomes a reality in a short timespan. Being in short supply, quantum-secure approaches to this threat will become in great demand.

Considering the selected papers for this study, the following tendencies have been observed in the grand area of quantum cryptography:

\subsection{Quantum Internet and networks}

Quantum Internet is an emerging technology in computer networks. Instead of sending bits from point-to-point, this technology sends, computes, and receives quantum states in the form of qubits. But the practical implementation of this technology still faces various challenges. 

One of the challenges is the limited distance that an optical fiber cable can cover~\cite{b1,b7}. A novel approach to quantum communication is the use of satellites, specially low-orbit satellites. The use of satellite-based communication showed a better end-to-end distance~\cite{b7}, but other communication challenges persist, such as low rate versus distance performance, a still lack of optimal codes for error correction~\cite{b12}, unintentional polarization shifts, the need for a special communication channel, and the difficult to implement multiplexing~\cite{b1}. In addition to the implementation of a communication channel between two ends with quantum capability, it is also important to think about the security of network domains without the possibility of optical connectivity for all end users~\cite{b11}.

According to~\cite{b51}, DV protocols with single-photon detection require specially designed systems and usually show a lower efficiency in comparison to CV systems applying weak coherent light. The paper also proclaims CV protocols can produce better key rates than DV protocols, but says DV protocols are simpler and with an unconditionally proven security.~\cite{b51} concludes saying that, for these reasons, it is not possible to tell whether DV or CV is better, and predicts that both approaches will continue to be pursued.

Man-in-the-Middle (MitM) attacks are easily detected in a quantum channel, but this property also makes Denial of Service (DoS) easier in a variety of protocols. With this in mind,~\cite{b25} concludes that availability should be guaranteed through backup solutions, and tools should be made to tell if the quantum channel was compromised due to the activity of an eavesdropper or due to noise in the channel.

\subsection{Quantum key distribution}

QKD have received attention from industry~\cite{b19}, but practical and efficient QKD protocols are necessary for the wide adoption and for the wide adoption of quantum cryptography in general. However, more research is still needed~\cite{b21}. This study mapped the following use cases for QKD:

\begin{itemize}
    \item In healthcare and on the Internet of Medical Things (IoMT)~\cite{b1,b9,b26,b41}.
    \item For mobile devices, such as the Internet of Things (IoT) and 5g/6g/7g technologies~\cite{b1,b11,b16,b21,b26,b29}.
    \item Unmanned Aerial Vehicle (UAV)~\cite{b3,b27,b31}.
    \item Research and development~\cite{b9}.
    \item Space and modern applications~\cite{b9}.
    \item Critical infrastructure and government~\cite{b9}.
    \item Internet banking~\cite{b9}.
    \item In the context of cloud computing and data centers~\cite{b26,b27,b31}.
    \item E-Commerce~\cite{b27}.
    \item Encrypted videoconferences~\cite{b27}.
\end{itemize}

And the following gaps and threats in QKD:

\begin{itemize}
    \item Improving distance limitations is one of the main current goals in QKD~\cite{b1,b11,b29}. Distance is one of the main factors in QKD, possibly affecting the loss of photons and leading to errors in the system~\cite{b9}.
    \item Just like improving distance limitations, QKD suffers from low key rates, and this is also one of the current main goals in the field~\cite{b1,b9,b11,b29,b32}.
    \item The elevated cost to implement QKD is still a relevant problem~\cite{b9,b29}. The cost of optical components makes the implementation prohibitively expensive~\cite{b1}.
    \item QKD can be affected by DoS and side-channel attacks~\cite{b1,b19,b49}.
    \item Single-photon sources are harder to construct than weak laser sources for QKD devices~\cite{b1}.
    \item Consumers have to feel confident about the products they are buying, argues~\cite{b1}. And complement saying that before implementing QKD networks, a better understanding of the weaknesses should be achieved and the lack of security standards is a concern.
    \item \cite{b9} tells us that QKD is prone to errors due to misalignment of optical components, noise in quantum detectors, disturbances in the quantum channel, or eavesdropping attempts. And concludes a key reconciliation is necessary in order to eliminate errors. In the same text, the authors argue the implementation of QKD requires expensive quantum devices (single-photon sources and detectors) that are currently considered inefficient, affected by noise, and operationally complex. The qubits require proper storage and management, otherwise they become vulnerable to eavesdropping. Hence, they conclude advanced hardware support and amplify qubits are required for implementation and key management.
    \item About scalability,~\cite{b9} argues quantum repeaters are necessary, because qubits lose information with time and repeaters should retain the energy stored to extend QKD over longer distances. However, retaining information from qubits after the decay in energy is still a very challenging task. They also state that the integration of quantum work with classical communication networks is very challenging while quantum information is still in its early stages of development.
    \item Technical and engineering flaws make QKD protocols vulnerable to attacks~\cite{b9}.
    \item \cite{b10} states that most quantum hacking strategies are directed at the single-photon detectors, being this the weakest point of QKD. And adds that the technological development of better detectors can also improve adjacent areas, such as quantum metrology and sensing.
    \item \cite{b11} suggests that, when or if a PQC solution is compromised, identifying a solution and replacing it would take a long time. And for this reason, one can trust only partially in PQC solutions, considering quantum computers could enable attacks beyond what is considered today. Other than that, PQC is intended to be secure against only known quantum attacks, whilst QKD is unconditionally secure. However, as a key agreement primitive, QKD is not intended to provide authentication, digital certificates, or digital signatures. So he defends a combination of PQC and QKD techniques to improve the security of the systems.~\cite{b19} also mentions the Internet in the future will probably be a hybrid quantum-classical Internet, making both PQC and QKD important subjects of study.
    \item Chip-based technology allows the implementation of QKD in mobile devices, having the benefits of low cost, low energy consumption, and compactness~\cite{b11}. But further advancements in cryptographic solutions for small devices should be made. Research should focus on improving both PQC and QKD technologies~\cite{b21}.
    \item According to~\cite{b18}, discrete modulation protocols are a strong alternative do gaussian modulation and can potentially benefit CV-QKD for longer distances. Larger data block sizes can also extend transmission distances and improve key generation rates, improving general security, with the trade-off of increasing computational demands.
    \item Artificial Intelligence (AI) can be used to aid in QKD algorithms, noise filtering, parameter optimization, system estimation, margin reduction, failure prediction, network self-configuration, dealing with quantum hacking, and in other aspects~\cite{b20,b31}. But further analysis of the applicability of Machine Learning (ML) algorithms to different protocols and channel types was not extensively done~\cite{b20}.
    \item The practical implementation of quantum cryptography requires scalability and integration with existing telecommunication infrastructure~\cite{b21}.
    \item Besides the theoretical analysis of quantum cryptography protocols, more practical experiments are necessary to validate the different protocols and countermeasures in real-world demonstrations~\cite{b21}.
    \item The potential of quantum hacking increases at the same pace quantum computing technology increases. Researchers should continue identifying potential security vulnerabilities in quantum cryptography~\cite{b21}.
    \item \cite{b26,b31} state that both DV-QKD and CV-QKD will very likely use the current optical fiber telecommunication infrastructure for classical transmission. This is due to a reduced total deployment cost. For this reason,~\cite{b26} says it is important to design efficient wavelength multiplexers with the capability of operating with both classical and quantum data transmission. He adds that the classical high power optical pulses can act as a noise source to the weak coherent pulses in QKD. And, in addition, concludes the technology for single-photon sources, single-photon detectors, and quantum repeaters should be improved.
    \item \cite{b27} argues that BB84 and other protocols use a classical channel for authentication and, for this reason, are prone to impersonation attacks.
    \item Currently, most discussions concerning QKD are limited to point-to-point communication, but networks are also a relevant topic~\cite{b29,b31}.
    \item \cite{b32} states a CV version of the Twin-Field protocol is absent, and also states a simplification and application of the Twin-Filed protocol in quantum communication protocols can promote practical applications.
    \item \cite{b35} addresses examples of what should be perfected in satellite-based QKD: channel transmissivity measurement, to understand the channel conditions; error correction, and the interface of satellite-based QKD with classical terrestrial networks.
    \item \cite{b49} addresses some quantum hacking concerns and vulnerabilities in QKD. A passive listening to the network is not possible, but a side-channel attack is possible, with the attacker exploiting imperfections without modifying characteristics of the implementation. On the other hand, in an active side-channel attack, the attacker does modify characteristics of the implementation. In a physical side-channel attack, the attack does not aim the theoretical foundations of a protocol, but exploits imperfections in unconventional channels such as: electromagnetic radiation, heat dissipation, acoustic noise, observation of computation time or power consumption.
\end{itemize}

\subsection{Quantum secure direct communications}

QSDC is a branch of quantum communication that aims at the direct transmission of messages, \textit{id est}, without the need for a security key.~\cite{b2} tells both QKD and QSDC were more extensively studied before 2015, but~\cite{b12} brought some open challenges in QSDC:
\begin{itemize}
    \item Most studies are focused on DV QSDC, but CV schemes are potentially more compatible with current telecommunication technology, leverage the advantage of cost-effective detectors, and potential higher rates.
    \item Designing CV-QSDC protocols that allow the receiver to detect information without the assistance of additional communication. One-way QSDC protocols would reduce channel losses and system complexity.
    \item More security proof of QSDC against quantum hackers should be done. And the same applies to the security analysis of imperfect devices.
    \item Improving performance and parameters of practical QSDC systems is a relevant topic for research.
    \item Considering that QSDC requires two-way block based transmission, the communication distance of single-photon QSDC would be half of QKD's for a certain rate. Combining quantum memory with QSDC is important to increase communication distance.
    \item Free-space optical QSDC is important for the future implementation of QSDC networks.
    \item It is possible to integrate QSDC into classical networks, but further efforts are needed.
\end{itemize}

\subsection{Quantum money/currency}

The concept of quantum money or quantum currency was already introduced in Section 4. The initial idea of a banknote resistant to forgery, leveraging quantum principles such as the no-cloning theorem, was first introduced by Stephen Wiesner in the 1970s, but was just published in~\cite{b63}.

Work still has to be done in the context of quantum cryptocurrency.~\cite{b21} says research should focus on new protocols and techniques that protect the user's privacy while maintaining the security of transactions.~\cite{b45} asks if quantum money can be built on standard cryptographic assumptions.

A list of open challenges in quantum money was brought by~\cite{b22}:
\begin{itemize}
    \item The secret key necessary to authenticate quantum money is encoded in quantum states and, hence, is vulnerable to external noise. Efficient key distribution is still an open challenge.
    \item Quantum computers and quantum money share technical constraints. At this point, quantum computers are still in their infancy and are currently unable to fully implement the algorithms required for quantum money.
    \item The high cost to implement quantum computer technology is still very high.
    \item A variety of vulnerabilities have to be resolved before implementing quantum money, such as eavesdropping and MitM attacks.
\end{itemize}

\subsection{Other challenges in quantum cryptography}

A diversity of other open challenges and trends in quantum cryptography have been found.~\cite{b4} pointed out that the keywords quantum blind signature, quantum cryptography, satellite-based QKD, and adiabatic quantum computing are gaining huge popularity lately.~\cite{b4} also pointed out that the fields of smart medical, smart industry, smart academia, and smart agriculture are paying more attention to PQC and quantum cryptography.

Below is a list of other open challenges in quantum cryptography:
\begin{itemize}
    \item As~\cite{b1} argues, the intricate rules of quantum mechanics hinder the advancement of quantum cryptography, making a foundation in physics crucial for both computer science researchers and users of quantum cryptographic products.
    \item According to~\cite{b1}, research should move towards QKD protocols, quantum repeaters, PQC, quantum network architectures, quantum entanglement, quantum hacking and countermeasures, quantum cryptographic protocols for new technologies, quantum communication in space, quantum cryptography standards, quantum-secured multi-party computation, quantum cryptographic hardware, and quantum cryptography in cloud computing.
    \item Known, unknown, and a combination of known and unknown types of quantum hacking should be considered~\cite{b1}.
    \item Quantum Internet, key distance, implementation cost, and preparing IoT for the quantum world will move at a rapid pace in the next few years~\cite{b4}.
    \item To perform key exchange under noisy conditions, maintaining key rates; entanglement-based QKD; quantum teleportation; quantum Internet; and QSDC. These are practical challenges in quantum communication protocols according to~\cite{b9}.
    \item Advancements in quantum hardware, especially error correction, and the development of scalable quantum processors are essential for the advancement of quantum cryptography~\cite{b13}.
    \item AI / ML has been used in QKD, quantum teleportation, QSS, and quantum networks~\cite{b14}.
    \item Quantum cryptography also imposes a threat to image encryption. Research efforts should be made to develop quantum resistant image encryption techniques, exploring PQC, QKD for secure transmission, and exploring the combination of classical and quantum approaches to image encryption~\cite{b13}. The secure exchange of multimedia data is another important topic~\cite{b44}.
    \item Quantum cryptography in voting security~\cite{b27}.
    \item Quantum cryptography in smart cards~\cite{b27}.
    \item Many open research topics on Quantum Oblivious Transfer (QOT)~\cite{b30}.
    \item An interesting question in semi-quantum cryptography is to explore "how quantum a protocol has to be to gain an advantage over its classical counterpart". Another interesting question is "how far one can go in reducing resource requirements and how this affects security". On the experimental side, an interesting question is to understand "what systems can be built and how"~\cite{b34}.
    \item Many unknown types of attack can come into existence in the future. In general, cryptography is geared towards commercial applications, and the efficiency of future protocols should be considered~\cite{b37}.
    \item A gap between perfect theory and imperfect practice generally exists in quantum cryptography, not only in QKD. When implementing or designing quantum cryptographic protocols, it is important to consider countermeasures and fault-tolerance to imperfections~\cite{b42}.
    \item Not only the theory, but also the practicability of quantum cryptography is an important challenge~\cite{b43}.
    \item The development of a device-independent protocol that tolerates a realistic amount of noise is an open challenge~\cite{b45}.
\end{itemize}

\section{Conclusion}

This tertiary review analyzed 51 secondary studies in the field of quantum cryptography available for consultation in the Scopus database. The aim of this review was to i) discover existing secondary studies in the field, exposing relevant bibliometric information and quality score; ii) understand the main quantum cryptographic techniques mentioned in these studies; iii) find tendencies, open challenges, or gaps in the field of quantum cryptography.

The results showed a low number of systematic studies in the field, with most authors adopting the form of a survey in their secondary studies($\approx 84.314\%$). On the other hand, the general quality of the selected papers was considered good, because most of the selected papers were classified as strongly correlated to the field of quantum cryptography ($\approx 94.118\%$) and the median year of publication was 2022.

This review found 15 keywords in quantum cryptography techniques to be relevant, with QKD prevailing untouched and mentioned in every selected study. Within it, 12 QKD protocols were mapped as relevant, with BB84 being both the oldest and the most popular to this date.

Being still in the vanguard of Computer Science, quantum cryptography still possesses plenty of open challenges. In this systematic review, a large number of open challenges and tendencies were exposed. In a few words, the attention to quantum cryptography has increased in the last three decades, but it still faces many problems. The cost of implementation is very high; the hardware still faces problems with decoherence, error correction, short communication distance, and low key rates. Most authors are concerned with point-to-point communication, but the implementation of quantum networks or quantum Internet is a bigger problem. In this area, scalability and device technology are crucial. Quantum cryptography is considered unconditionally secure in theory, but practical implementations can suffer side-channel or DoS attacks. These threats have to be resolved in order to produce trustworthy quantum communication, be it with QKD, QSDC, quantum money, or other quantum technologies. However, quantum cryptography has a high chance of becoming commercially relevant in the near future.

\section{Funding}
This work was supported by the Fundacao de Amparo a Pesquisa e Inovacao do Estado de Santa Catarina (FAPESC), Edital 62/2024.

\end{document}